# Observation of superluminal signaling of terahertz pulses

Liang Wu[1,2], Zhiyong Wang[1]*, Kai Kang[1], Yi Fu[2], Chuanwei Li[1], Weili Zhang[3]* and Shuang Zhang[4]*

[1]Mechanics Department, School of Mechanical Engineering, Tianjin University, Tianjin 300072, China.

[2]School of Precision Instrument and Optoelectronics Engineering, Tianjin University and the Key Laboratory of Optoelectronics Information and Technology (Ministry of Education), Tianjin 300072, China.

[3]School of Electrical and Computer Engineering, Oklahoma State University, Stillwater, Oklahoma 74078, USA.

[4]School of Physics and Astronomy, University of Birmingham, Birmingham B15 2TT, United Kingdom.

*Correspondence to: zywang@tju.edu.cn, weili.zhang@okstate.edu, s.zhang@bham.ac.uk.

**Abstract:** Superluminal tunneling of light through a barrier has attracted broad interest in the last several decades. Despite the observation of such phenomena in various systems, it has been under intensive debate whether the transmitted light truly carry the information of the original pulse. Here we report observation of anomalous time response for terahertz electromagnetic pulses passing through thin metal films, with the pulse shape of the transmitted beam faithfully resembling that of the incident beam. A causal theoretical analysis is developed to explain the experiments, though the theory of Special Relativity may confront a challenge in this exceptional circumstance. These findings may facilitate future applications in high-speed optical communication or signal transmission, and may reshape our fundamental understanding about the tunneling of light.

Ever since the discovery of quantum mechanical tunneling, the time needed for a wave packet to pass through a barrier has attracted the interests of researchers for decades[1]. L.A. MacColl[2] performed the first theoretical analysis for the dynamics of tunneling process and suggested that there is "no appreciable delay in the transmission of the packet through the barrier". A later investigation yielded that the finite traversal phase time is independent of medium width and would saturate with distance, which is known as the "Hartman effect" in literature[3]. These works have initiated a longstanding paradox for physicists.

In the past decades, triggered by the progress in device processing and micro- and nano-technology, and assisted by a deepened understanding of the origin of optical tunneling, tremendous effort has been dedicated to measuring the tunneling time of classical wave packets via non-invasive methods[4-7]. Spielmann et al. reported the first experimental observation of tunneling of optical pulses through photonic band gap barriers[5]. S. Longhi et al. performed time-resolved measurement of superluminal tunneling of picosecond optical pulses in a periodic fiber Bragg grating, and a propagating group velocity twice the speed of light in vacuum was claimed[6]. The experimental investigation of the tunneling time has also been extended to microwave, radio, terahertz, or even acoustic frequencies[8-15]. For example, J. J. Carey et al. reported the observation of non-causal tunneling time response with single-cycle terahertz pulses in frustrated total internal reflection, and claimed that both the phase and energy of the pulse travel faster than the speed of light in vacuum[12]. M.T. Reiten et al. carried out a similar time-domain measurement of tunneling of terahertz electromagnetic pulses through an air gap between two prisms, and interpreted their results based on linear dispersion without considering the ray trajectory through the barrier[13]. They concluded that there was no acausal effect in the tunneling process.

In order to understand the dynamics and physical origin of the tunneling process, several assumptions and theories were suggested, leading to a heated debate for many years. A prevalent opinion is that the transmitted wave packet is merely reshaped[4-6], without violating the causality. On the other hand, H. G. Winful[16,17] proposed an alternative interpretation of the anomalously short delay times of the tunneling electromagnetic pulses, based on the argument that this is a quasistatic process consisting of energy storage and its subsequent release of standing waves. This explanation seems to have solved the longstanding debate. Most recently, H. Y. Yao et al. developed a time-domain framework to analyze the origin of faster-than-light phenomenon, which seems to arise from the destructive interference of multiple-reflection echoes[18]. Hence, *a consensus has been reached that superluminal information transmission is not possible*, based on the argument that the majority of the incident wave-packet energy is reflected, and there is no proof that the transmitted particle or wave packet is the same as the incident one before tunneling[19].

Here we experimentally demonstrate anomalous terahertz phase transmission through metal films of different thicknesses via time domain measurements. To confirm that the transmitted signals carry the same spectral features as the incident pulses, a terahertz metasurface exhibiting Fano resonant spectral feature is exploited to modulate the incident terahertz waves. Contrary to the widely accepted belief, the experimental results here, along with numerical calculations, serve as direct evidence of anomalous phase index and superluminal signaling for terahertz pulses transmitting through the metal films. Because there is no definite beam path or trajectory through the metal film, namely, a group velocity cannot be well defined. Hence, we name the observed phenomenon as "superluminal signaling". Our work opens up new prospects for studies of superluminal tunneling effects and it enables potential applications in ultra-high speed communication in the technologically difficult terahertz-frequency regime.

The samples are prepared by depositing metal films on a 300-μm-thick high-resistivity single-crystal silicon. Cu film and Ag film are deposited by electron beam evaporation with a vacuum condition of 2 x $10^{-6}$ Pa, and an evaporation rate of 1.5 Å/s to ensure the quality of the film. For each metal, samples with thickness of 10 nm and 20 nm are prepared. A Bruker DektakXT step profiler is used to measure the thickness of the metal film, which shows a fluctuation level of ±2 nm. In order to stabilize the surface composition, the targets are cleaned with acetone and alcohol. All the films are deposited on Si substrates at room temperature.

In our experiment, the transmitted time-resolved electric fields of normal incident terahertz wave packet are measured with femtosecond accuracy using a typical 8f confocal terahertz time-domain spectrometer (THz-TDS)[20]. The terahertz signals passing through an uncovered area of the same sample pieces are used as a reference. A metasurface fabricated by conventional photolithography process is employed in the measurement to modulate the incident terahertz pulses. The metasurface consists of asymmetric split rings with two arms of different lengths, with the detail structural parameters provided in supplementary information. In order to improve the precision, we repeat all the measurements 5 times and averaged the testing values. The experimental configuration and the used signal processing algorithm are presented in supplementary information.

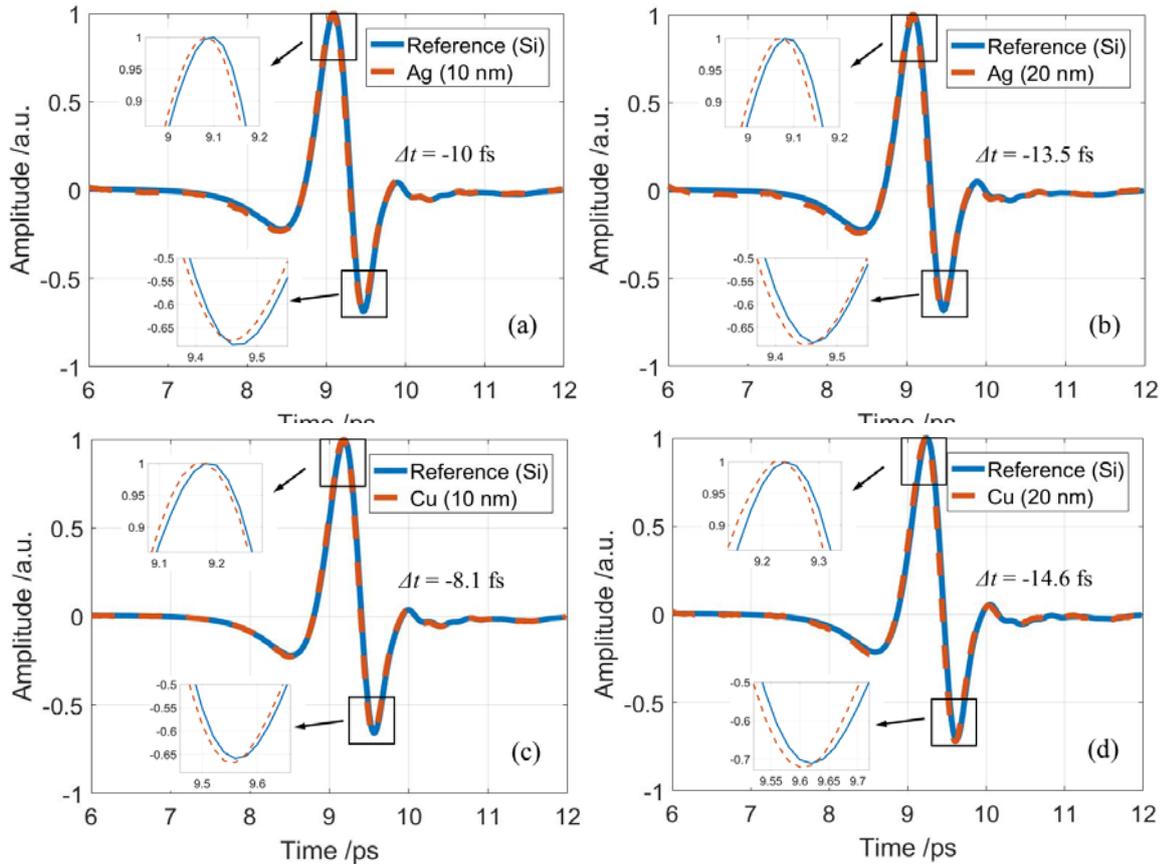

**Fig. 1| Measured terahertz pulses transmitted through metal films of a,** 10-nm Ag; **b**, 20-nm Ag; **c,** 10-nm Cu; **d,** 20-nm Cu.

Figure 1 shows the experimental results of the terahertz pulses propagating through different metal films. It is observed that the wave forms of the pulses after transmitting through the metal films are nearly identical to that through the bare silicon wafer. By zooming in at the pulse peak, it is found that the pulses transmitting through the samples arrive earlier than that through the silicon substrate. The time difference between the reference (bare Si wafer) and the samples is -10 fs, -13.5 fs, -8.1 fs, -14.6 fs for the four films, respectively, showing that all the tunneled wave packets arrive earlier than that through the bare silicon wafer. Thus, we demonstrate that the anomalous time advance is a robust phenomenon, with the measurement details provided in supplementary information.

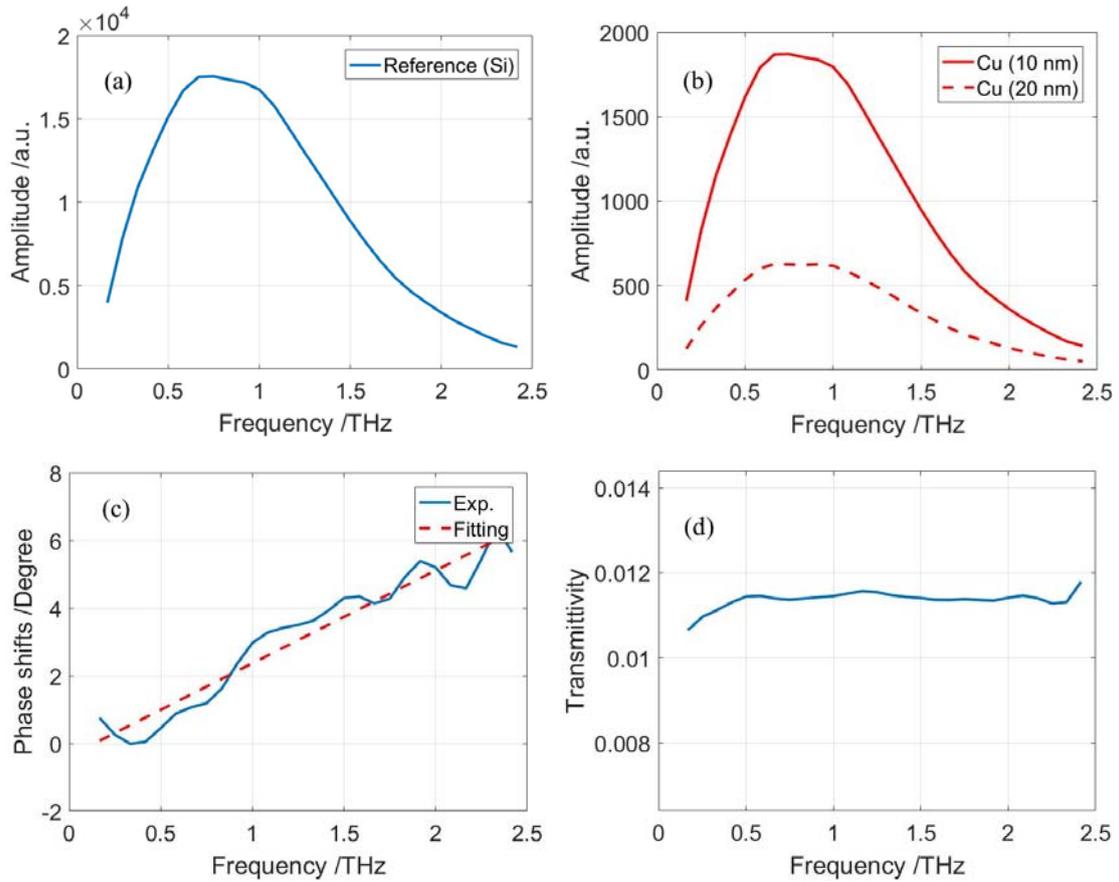

**Fig. 2| Measured amplitude spectra corresponding to the terahertz pulses in Fig.1: a,** reference (Si); **b,** Cu films. **c,** Frequency dependence of phase shift for the transmitted terahertz pulses through 10-nm Cu film. **d,** Transmittivity of 10-nm Cu.

We further perform a Fourier transform and obtain the amplitude and phase of the transmission spectra for bare silicon and for Cu/silicon, which are shown in Figures 2a and 2b, respectively. Figures 2c and 2d show the frequency dependence of the phase shift and the amplitude for the transmitted pulses through a 10-nm Cu film in the frequency range of 0.2-2.4 THz, respectively. A nearly constant transmission is obtained within this frequency range. Despite the roughness of the phase profile, a linear trend with a positive slope is clearly observed which also indicates the advance in time.

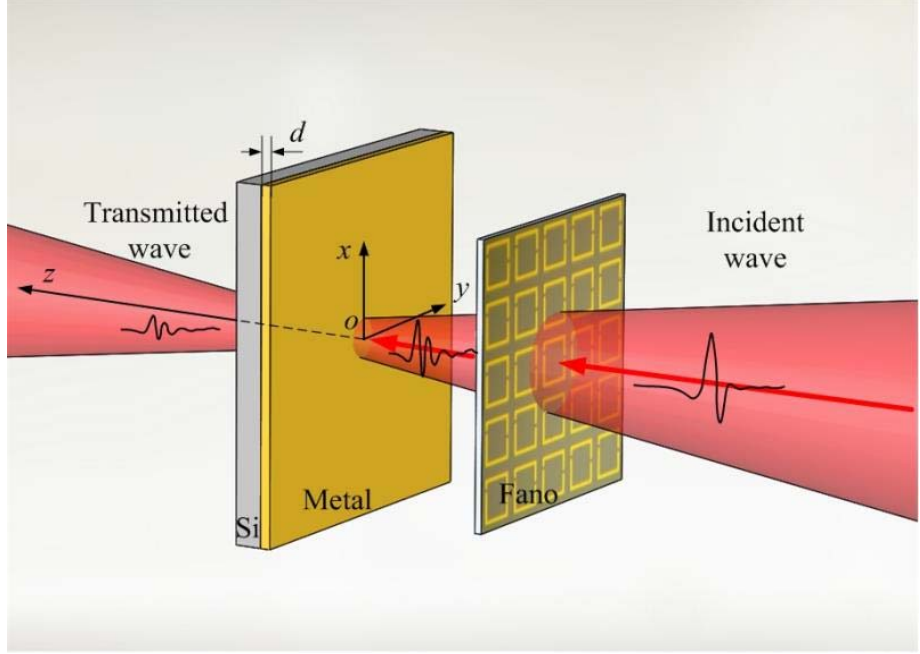

**Fig. 3| Diagram of experimental design. Incident terahertz waves transmit through a metal film medium of width d.** A Fano metastructure is utilized in the subsequent measurement.

The superluminal effect is analyzed using transmission theory. Consider a monochromatic wave with frequency $\omega$ incident normally onto a metal film medium and let $\varepsilon(\omega)$, $\mu(\omega)$ be the frequency-dependent material parameters. For simplicity, we consider the substrate to be the same medium as air with index $n_0=1$. Note that this does not affect the calculating results (detail mathematical proof can be seen in supplementary information). The transmittance can be obtained by using the boundary conditions at each interface[21],

$$T(\omega)=|T(\omega)|e^{i\phi\omega} = \frac{1}{\cos(qd)-\frac{i}{2}(\frac{q}{\mu k}+\frac{\mu k}{q})\sin(qd)}, \tag{1}$$

where $k=\omega/c$, $q=\omega/c(\varepsilon(\omega)\mu(\omega))^{1/2}$ are wave vectors in air and metal slab, respectively. The transmitted pulse could be written as the integral of each plane-wave component propagated through the metal film,

$$E_T(z,t)=\int_{-\infty}^{\infty}d\omega A(\omega)|T(\omega)|e^{i\phi(\omega)-i\omega t}e^{\frac{i\omega}{c}(z-d)}, \tag{2}$$

where $A(\omega)$ denotes the amplitude spectrum of the incident wave. To evaluate the delay time of the peak of the transmitted wave-packet, one can take the first-order Taylor series expansion about the carrier frequency, then the tunneled phase time can be defined as

$$t_\phi = \frac{\partial \phi(\omega)}{\partial \omega}\bigg|_{\omega_c}, \tag{3}$$

where $\varphi(\omega)$ is the phase of the transmission amplitude, and $\omega_c$ is the carrier frequency of the pulse.

According to the Drude model, the dielectric function of metals is written as

$$\varepsilon(\omega)=1-\frac{\omega_p^2}{\omega(\omega+i\gamma)}, \qquad (4)$$

where $\omega_p=\sqrt{4\pi n_B e^2/m^*}$ is the bulk plasmon frequency, with $n_B$ the bulk electron density and $m^*$ the effective electron mass, $\gamma$ is a measure of the inelastic scattering rate, which is strongly dependent on the film thickness. At terahertz frequencies which are far below the plasma frequency of metals, the permittivity is very negative, i.e. $\varepsilon(\omega)<0$, $\mu(\omega)>0$. The phase time $t_\varphi$ calculated from the Eq. 3 confirms that superluminal delay times will be obtained. (see details in supplementary information or in Reference[21])

It can be seen from Fig. 2d that the transmittivity of 10-nm Cu is only ~1%, that is, ~99% of the incident energy is reflected or absorbed during the tunneling process, so one cannot be certain that the wave packet that transmits through the sample is the same wave packet that is incident onto the sample. In order to solve this dilemma, we utilize a Fano metasurface at the incident face to manipulate the pulse, and detect the modulated tunneling signal, as shown in Figure 3. For this purpose, a 20-nm Cu film is selected for the measurement with a transmittivity around 0.3%.

The transmitted waveforms in time domain are measured and shown in Figure 4a. The transit time of 20-nm Cu compared to air is -14.5 fs, which is in excellent agreement with the measurement result without Fano resonator. The corresponding phase spectrum is shown in Figure 4b. A linear fit with positve slope in frequency range represents the time advance.

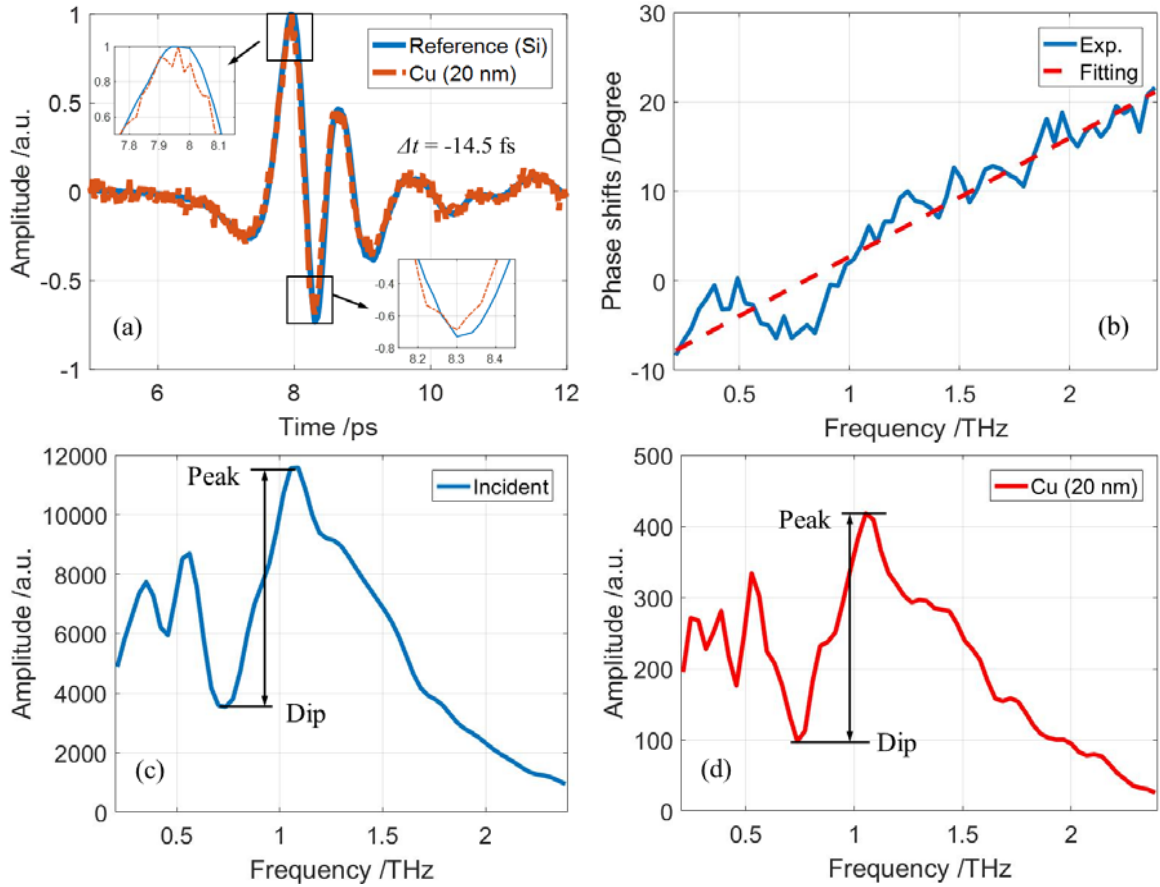

**Fig. 4| a,** Measured Fano pulses transmitted through 20-nm Cu film. The reference is also plotted. The detail signal processing algorithm is presented in supplementary information. **b,** Phase spectrum of incident Fano modulated terahertz pulses passing through 20nm-Cu specimen. **c,** Measured amplitude spectra of the Fano pulses propagating through reference (Si) and **d,** 20nm-Cu specimen.

Measured amplitude spectra of the Fano pulses calculated by Fourier transform are displayed in Figures 4c and 4d. The commonly used Fano resonance is around 0.5-0.6 THz frequency, corresponding to a high Q mode. A more distinguishable subsequent dip here, arising from a dipole oscillation mode, is perceived as the main dip. It is noted that, different from the others, the amplitude peak at about 1.1 THz is not an oscillation mode, but a superposition of the main dip and the natural decrease of the source frequency amplitude, as shown in Figs. 2a and 4c. Nevertheless, this coincident "main peak" becomes another characteristic of the incident pulse. It is obvious in Fig. 4d that the key features of the incident pulse have traveled to the exit albeit the large transmission loss, and we can identify this special signal from a common one in frequency domain. If we define the original terahertz signal as "0", and Fano modulated signal as "1", then superluminal information transmission cloud be achieved, establishing the potential for ultra-high-speed communication in future.

It is worthy to note that the Fano vibration strength could be tuned by external fields, or we can set a coding metamaterial[22] at the input, either of which can raise the amount of data transmitted. Moreover, if one use careful-designed optical barriers instead of metal films, for example, a photonic bandgap or a frustrated total internal reflection structure, much faster signal transmission would also be predicted.

**Conclusion**

We have measured the amplitude and phase change of terahertz pulses by THz-TDS following transmission through the metal films. There are two findings observed in the experiments:

1. The transmitted pulse arrives earlier in time than that past the same space in vacuum;

2. The substantial features of the incident pulse are remained in the transmitted pulse.

No specific passage of the energy propagation could be assigned here. However, no matter the tunneled pulse is merely reshaped or the release of energy storage or the superposition of multiple internal reflections, it indeed contains information from the input. Whether causality still works in this case or not, is a question.

**References**

1. Condon, E.U. Quantum mechanics of collision processes. *Rev. Mod. Phys.* **4**, 577 (1931).

2. MacColl, L. A. Note on the transmission and reflection of wave packets by potential barriers. *Phys. Rev.* **40**, 621-626 (1932).

3. Thomas, E. H. Tunneling of a wave packet. *J. Appl. Phys.* **33**, 3427 (1962).

4. Steinberg, A.M. et al. Measurement of the single-photon tunneling time. *Phys. Rev. Lett.* **71**, 708 (1993).

5. Spielmann, Ch. et al. Tunneling of optical pulses through photonic band gaps. *Phys. Rev. Lett.* **73**, 2308 (1994).

6. Longhi, S. et al. Belmonte, Superluminal optical pulse propagation at 1.5 μm in periodic fiber Bragg gratings. *Phys. Rev. E* **64**, 055602 (2001).

7. Balcou, P. et al. Dual optical tunneling times in frustrated total internal reflection. *Phys. Rev. Lett.* **78**, 851-854 (1997).

8. Ranfagni, A. et al. Delay-time measurements in narrowed waveguides as a test of tunneling. *Appl. Phys. Lett.* **58**, 774-776 (1991).

9. Enders, A. & Nimtz, G. Evanescent-mode propagation and quantum tunneling. *Phys. Rev. E* **48**, 632 (1993).

10. Mojahedi, M. et al. Abnormal wave propagation in passive media. *IEEE J. Sel. Top. Quant. Electron.* **9**, 30-39 (2003).

11. Hache, A. & Poirier, L. Long-range superluminal pulse propagation in a coaxial photonic crystal. *Appl. Phys. Lett.* **80**, 518-520 (2002).

12. Carey, J.J. et al. Noncausal time response in frustrated total internal reflection?. *Phys. Rev. Lett.* **84**, 1431-1334 (2000).

13. Reiten, M.T. et al. Optical tunneling of single-cycle terahertz bandwidth pulses. *Phys. Rev. E* **64**, 036604 (2001).

14. Yang, S. et al. Ultrasound tunneling through 3D phononic crystals. *Phys. Rev. Lett.* **88**, 104301 (2002).

15. Robertson, W.M. et al. Breaking the sound barrier: tunneling of acoustic waves through the forbidden transmission region of a one-dimensional acoustic band gap array. *Am. J. Phys.* **70**, 689-693 (2002).

16. Winful, H.G. Energy storage in superluminal barrier tunneling: Origin of the "Hartman effect". *Opt. Express* **10**, 1491-1496 (2002)

17. Winful, H.G. Nature of ''Superluminal" Barrier Tunneling. *Phys. Rev. Lett.* **90**, 023901 (2003)
18. Yao, H. Y. & Chang, T. H. Time-domain analysis of superluminal effect for one-dimensional Fabry-Pérot cavity. *Chinese J. Phys.* **67**, 657–665 (2020).
19. Winful, H.G. Tunneling time, the Hartman effect, and superluminality: A proposed resolution of an old paradox. *Phys. Reports* **436**, 1-69 (2006)
20. Wu, L. et al. A new $Ba_{0.6}Sr_{0.4}TiO_3$-silicon hybrid metamaterial device in terahertz regime. *Small*. **12**, 2609 (2016).
21. Gupta, S. D. et al. Subluminal to superluminal propagation in a left-handed medium. *Phys. Rev. B* **69,** 113104 (2004).
22. Cui, T. J. et al. Coding Metamaterials, digital Metamaterials and Programming Metamaterials. *Light-Sci Appl.* **3**, 218 (2014).

## Acknowledgments

The authors acknowledge useful discussions with Prof. John O'Hara. The authors acknowledge financial support from the National Key Research and Development Program of China (2018YFB0703500), National Natural Science Foundation of China (11827802, 11772222).

## Authors contributions

L.Wu and Z.Y. Wang initiated the project. K. Kang and Y. Fu performed the experiments. C.W. Li developed the signal processing algorithm. Z.Y. Wang, Y. Fu. and K. Kang prepared all the figures. L.Wu developed the theoretical model and prepared the manuscript. W. L. Zhang. and S. Zhang supervised the project.

## Additional information

All data are available in the manuscript or the supplementary information.

## Competing financial interests

The authors declare no competing financial interests.

# Supplementary Information for

## Observation of superluminal signaling of terahertz pulses

Liang Wu, Zhiyong Wang, Kai Kang, Yi Fu, Chuanwei Li, Weili Zhang, and Shuang Zhang.

Correspondence to: zywang@tju.edu.cn, weili.zhang@okstate.edu, s.zhang@bham.ac.uk.

**This PDF file includes:**

    Materials and Methods
    Supplementary Text
    Figs. S1 to S5
    Table S1

**Materials and Methods**

**1. Materials**

Detail structural parameters of the utilized Fano metamaterial are shown in Figure S1. The asymmetric split ring arrays were patterned using conventional photolithography on high resistivity Si wafer (with a conductivity of 0.03 S/m), followed by 200-nm gold (Au) film deposition. In order to enhance adhesion between Au and Si, a 10 nm titanium (Ti) layer under Au was also required.

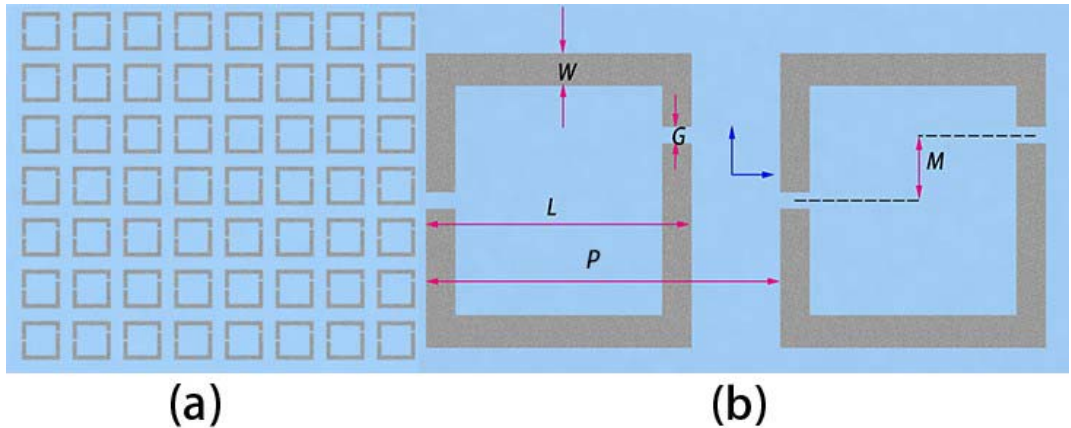

**Fig. S1| a,** Microscopic image of the fabricated Fano metamaterial. **b,** Schematic of the unit structure with parameters: M = 15 μm, G = 3 μm, W = 6 μm, L = 60 μm, P = 75 μm.

**2. More Details about Experimental Measurements**

**2.1 Experimental system and signal processing method**

Figure S2 illustrates the schematic of our used THz-TDS system. In the system, a femtosecond laser with 88 MHz repetition rate, 26 fs pulse width and 800 nm wavelength is used as the pump source with a 10 mw average power for both the terahertz transmitter and receiver. Terahertz radiation is generated and detected by shorting the dipole-type photoconductive antenna gap. To obtain high signal stability, the ambient humidity is kept below 1% by pumping dry air into a chamber surrounding the system.

The time shifts in Fig. 1 and Fig. 4 are obtained by different methods. The experimental schemes are shown in Fig. S2. There are two apertures on the used sample holder. In the experiments without Fano resonator, a Si wafer used as reference and metal film deposited on Si wafer cut off from the same piece are placed on the two apertures. After drying the chamber, the signals through Si wafer and metal film specimen are captured, as shown in Fig. 1. The two signals, marked as A and B in Fig. S2a, are captured after one time of drying the chamber. A signal matching algorithm to obtain the maximum correlation coefficient between the two signals is developed to calculate the shift between them. Hereafter, the used algorithm will be referred as matching algorithm.

In the experiments with Fano resonator, four signals are captured for one metal film specimen as shown in Fig. S2b. Four signals, marked as A, B, C and D, are captured after two times of drying the chamber to ensure that the same Fano resonator is used for Si reference and metal film specimen. Fig. S3 shows the captured four signals for the 20-nm Cu specimen. Because there may be a slight difference in the environmental condition between the two times of drying the chamber, it is not appropriate to calculate the times between the B and D signals. The used data processing flow is explained as follows. Firstly, the matching algorithm is used to calculate the time shift between A and C signals, with the obtained time shift notated as $t_{AC}$, which represents the time shift caused by environment change between the two times of drying the chamber. Then, signal D is numerically translated to eliminate this time shift caused by environment change. The waveforms in Fig. 4a are the original B signal and translated D signal. At last, the matching algorithm is used to calculate the time interval between the original B signal and the translated D signal.

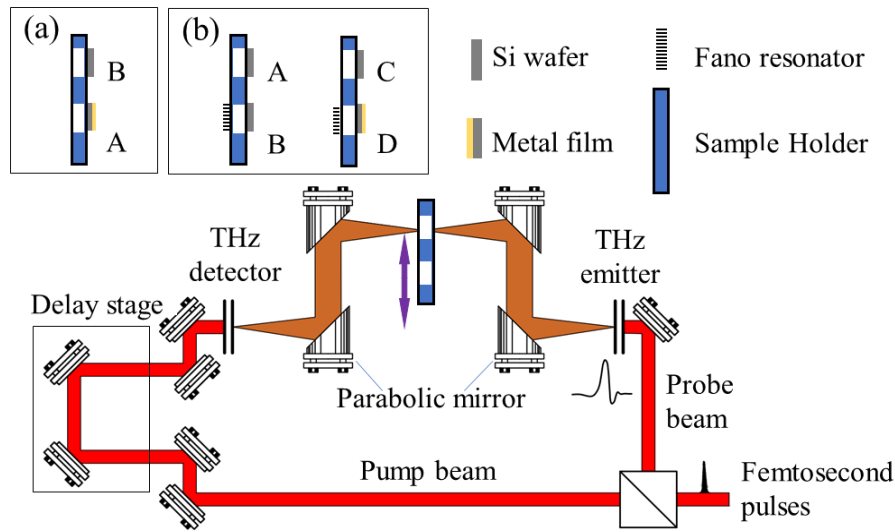

**Fig. S2| The used THz-TDS system. a,** The Scheme for experiments without Fano resonator; **b,** the Scheme for experiments with Fano resonator.

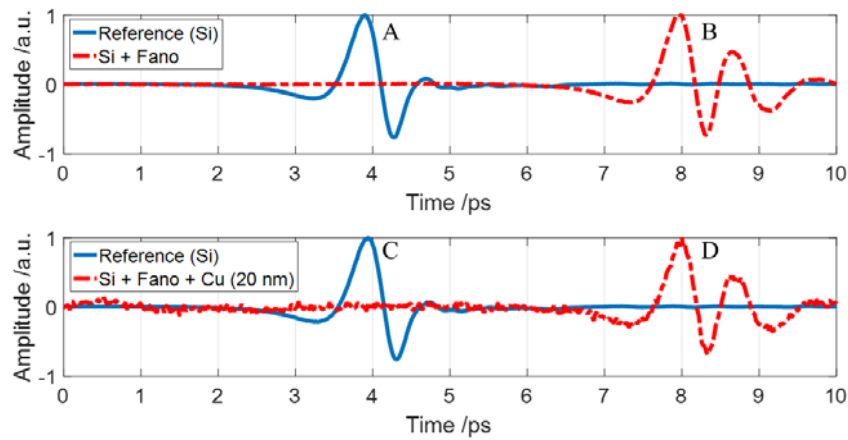

**Fig. S3| The captured four signals for the 20-nm Cu specimen.**

## 2.2 Original data

In our experiments, five times of signal captures are conducted repeatedly for every experimental situation. The signals presented in Fig. 1 and Fig. 4 are obtained by averaging the signals from the five capturing. The detail view of original signals around their peaks are presented in Fig. S4. In Fig. S4, there are some randomly shifts in the original signals, but the signals through metal-film specimen are roughly ahead of the reference signals through bare Si in time.

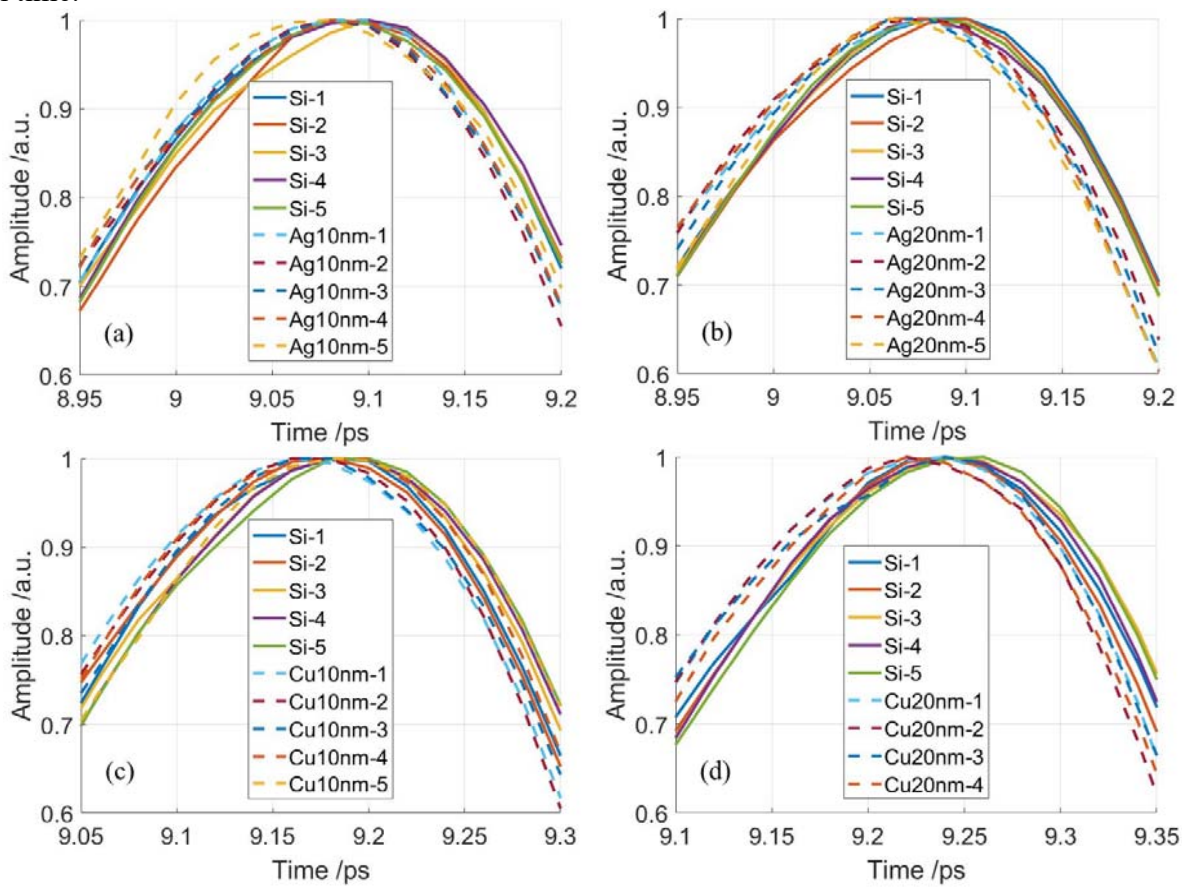

**Fig. S4| Original terahertz pulses transmitted through metal films of a,** 10-nm Ag; **b,** 20-nm Ag; **c,** 10-nm Cu; **d,** 20-nm Cu.

The following process is used to analyze the random shifts. Firstly, a mean signal is obtained by averaging the signals from the five capturing. The matching algorithm mentioned above is used to calculate the shift between the mean signal and each original signals. The results are listed in Table S1. For convenience of comparison, the standard variance of random shifts and the time shifts of metal film specimens relative to Si reference are also listed in Table S1. Except for the 10 nm Cu specimen, the standard variances of random shifts are much less than the time shifts of metal film specimens. Thus, even if the random shifts are considered, it is still a reliable conclusion that the signals through metal film specimens are ahead of the signals through bare Si wafer.

**Table S1| The random shifts of original signals and the delay time of metal film specimen**

|  |  | $\Delta t_r$ (fs) | | | | | $\sigma$ | $\Delta t$ (fs) |
|---|---|---|---|---|---|---|---|---|
| 10 nm Ag | Si | -3.3 | 2.1 | -0.2 | 1.7 | 0.2 | 2.1 | -10 |
|  | Si + 10 nm Ag | 1.9 | -2.8 | 0.7 | 0.4 | -0.2 | 1.7 |  |
| 20 nm Ag | Si | 2.2 | 0.8 | -2.2 | 1.1 | -1.8 | 1.9 | -13.5 |
|  | Si+ 20 nm Ag | 0.1 | 2.3 | 0.3 | -3.9 | 1.3 | 2.4 |  |
| 10 nm Cu | Si | -5.4 | -6.3 | 1.0 | 5.3 | 5.33 | 5.6 | -8.1 |
|  | Si + 10 nm Ag | -4.9 | 8.3 | -0.2 | 1.8 | 11.9 | 7.7 |  |
| 20 nm Cu | Si | -1.8 | -5.4 | 4.3 | -0.9 | 3.7 | 4.0 | -14.6 |
|  | Si + 20 nm Ag | -0.6 | -0.4 | 5.1 | 0 | none | 3.9 |  |
| $\Delta t_r$: the random shifts between the mean signal and original signals | | | | | | | | |
| $\sigma$: the standard variance of $\Delta t_{Random}$ | | | | | | | | |
| $\Delta t$: the time shift of metal film specimen relative to Si reference (negative sign means advance) | | | | | | | | |

## 2.3 An Experiment to determine the direction of time

To illustrate whether the signal shift to the left represents advance or delay in Fig. 1 and Fig. 4, an experiment has been conducted. We have captured three signals through air, bare Si wafer, and 20 nm Cu specimen. Figure S5 presented the captured signals. In Fig. S5, the pulse through Si wafer is on the right of that through air, thus the right shift represents delay. On the contrary, the left represents advance. In Fig. S5, the pulse through 20 nm Cu specimen is on the left of that through Si wafer, thus the shift between them means advance, instead of delay.

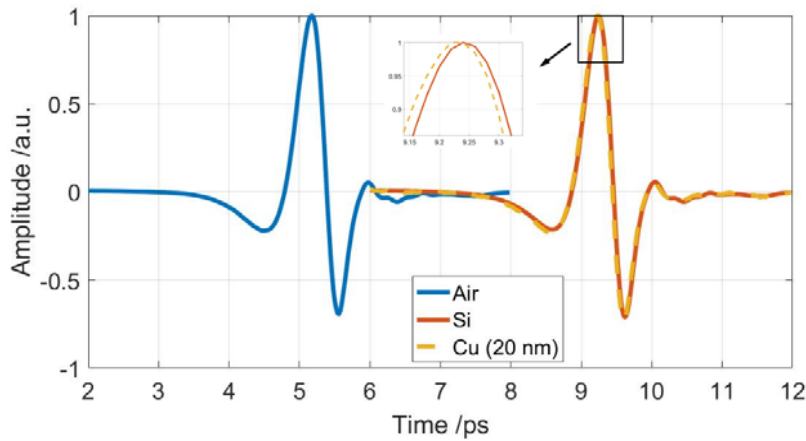

**Fig. S5| The pulses transmitted through air, Si wafer and 20 nm Cu specimen.**

## 2.4 Some details about the mechanical delay stage

A PI precision linear stage M406.2DG is used in our experiments. According to the specimen, its minimum incremental motion is 0.1μm (the light flight time during this distance is about 0.67 fs). The design resolution of the used stage is 0.0085μm, and an absolute encoder is used in the system. In our measurements, the stage returns to the same start position at the end of each scan, and all the scans move the same direction in each run.

**2.5 The relation between the phase in frequency domain and the delay in time domain**

Consider $G(f)$ is the Fourier transform of $g(t)$, i.e.

$$G(f) = \int_{-\infty}^{+\infty} g(t) e^{-i2\pi ft} dt, \tag{S1}$$

the Fourier transform of $g(t+\delta)$, noted as $G_\delta(f)$, should be

$$G_\delta(f) = \int_{-\infty}^{+\infty} g(t+\delta) e^{-i2\pi ft} dt = e^{i2\pi\delta f} G(f). \tag{S2}$$

If $\delta > 0$, $g(t+\delta)$ is advanced relative to $g(t)$. The phase difference between $G_\delta(f)$ and $G(f)$ should be $2\pi\delta f$, whose slope is positive. Thus, the positive slope of phase means the advance in time.

**Supplementary Text**

**Theoretical analysis details**

The plane-wave solutions for the electric-field vector $\vec{E}(z,t) = \vec{E}(z)e^{-i\omega t}$ in different regions are given by

$$\begin{aligned} E(z) &= Ae^{ikz} + Be^{-ikz}, (z \leq 0) \\ &= Ce^{iqz} + De^{-iqz}, (0 \leq z \leq d) \\ &= Fe^{ik(z-d)}, (z \geq d) \end{aligned} \tag{S3}$$

Here $k=\omega/c$, $q=\omega/c(\varepsilon(\omega)\mu(\omega))^{1/2}$ are wave vectors in air and metal film, respectively. Then Eq. 2-4 in the manuscript can be derived.

We perform differentiation of Eq. 4, and calculate the phase time result as follows

$$t_\phi = \frac{P}{D},$$

$$P = \frac{1}{2} \tan \frac{\omega d N(\omega)}{c} \left( \frac{1}{\mu} \frac{\partial N}{\partial \omega} - \frac{\mu}{N^2} \frac{\partial N}{\partial \omega} - \frac{N}{\mu^2} \frac{\partial \mu}{\partial \omega} + \frac{1}{N} \frac{\partial \mu}{\partial \omega} \right)$$

$$+ \frac{1}{2} \left( \frac{N}{\mu} + \frac{\mu}{N} \right) \sec^2 \frac{\omega d N(\omega)}{c} \frac{\partial}{\partial \omega} [\omega N(\omega)] \frac{d}{c}$$

$$D = 1 + \frac{1}{4} \left( \frac{N}{\mu} + \frac{\mu}{N} \right)^2 \tan^2 \frac{\omega d N(\omega)}{c}$$

$$\varepsilon(\omega)=1-\frac{\omega_p^2}{\omega^2}, \mu(\omega)=\frac{(\omega^2-\omega_m^2)}{(\omega^2-\omega_0^2)}$$

$$N(\omega) \equiv \sqrt{\varepsilon(\omega)\mu(\omega)} \tag{S4}$$

where $\omega_m$ is "magnetic plasma" frequency. In the terahertz frequency range, the term $\omega_m < \omega < \omega_p$ is matched, that is, $\varepsilon(\omega)<0$ and $\mu(\omega)>0$, so the superluminal tunneling times could understood from Eq. S4.

**Proof of assumption**

The substrate with index $n_1$ equals to air with index $n_0=1$ for the calculating results. From Maxwell's equations,

$$E_1 = Ae^{ikz} + Be^{-ikz}, H_1 = \frac{A}{\eta_1}e^{ikz} - \frac{B}{\eta_1}e^{-ikz} \ (z \le 0, \ \eta_1 = \frac{ck}{\omega} \ \mu_1=1) \tag{S5}$$

$$E_2 = Ce^{iqz} De^{iqz}, H_2 = \frac{C}{\eta_2}e^{iqz} - \frac{D}{\eta_2}e^{-iqz} \ (0 \le z \le d, \ \eta_2 = \frac{cq}{\mu\omega}, \ \mu_2=\mu) \tag{S6}$$

$$E_3 = Fe^{ik'(z-d)}, H_3 = \frac{F}{\eta_3}e^{ik'(z-d)} \ (z \ge d, \ \eta_3 = \frac{ck'}{\omega}, \ \mu_3=1) \tag{S7}$$

At the interface z=0,

$$E_{10} = A+B = C+D = E_{20}, H_{10} = \frac{A-B}{\eta_1} = \frac{C-D}{\eta_2} = H_{20} \tag{S8}$$

At the interface z=d,

$$E_{2d} = Ce^{iqd} + De^{iqd} = F = E_{3d}, H_{2d} = \frac{1}{\eta_2}\left(Ce^{iqd} - De^{iqd}\right) = \frac{F}{\eta_3} = H_{3d} \tag{S9}$$

According to Eq. S6,

$$A = \frac{1}{2}\left(\frac{\eta_1+\eta_2}{\eta_2}\frac{\eta_2+\eta_3}{2\eta_3}Fe^{-iqd} + \frac{\eta_2-\eta_1}{\eta_2}\frac{\eta_3-\eta_2}{2\eta_3}Fe^{iqd}\right) \tag{S10}$$

Substitute the expressions of $\eta_x$ from Eqs. S3-S5 into Eq. S8,

$$T = \frac{A}{F} = \left[\cos(qd) - \frac{i}{2}\left(\frac{q}{\mu k} + \frac{\mu k}{q}\right)\sin(qd)\right]^{-1} \tag{S11}$$

Proved.